\documentclass[prl,twocolumn,superscriptaddress,letterpaper]{revtex4}
\usepackage{graphicx,amssymb,bm,natbib,amsfonts,amsmath}
\usepackage{hyperref}

\begin{document}

\title{Many-body interactions in quasi-freestanding graphene}

\author{David A. Siegel}
\affiliation{Department of Physics, University of California,
Berkeley, CA 94720, USA}
\affiliation{Materials Sciences Division,
Lawrence Berkeley National Laboratory, Berkeley, CA 94720, USA}

\author{Cheol-Hwan Park}
\affiliation{Department of Physics, University of California,
Berkeley, CA 94720, USA}

\author{Choongyu Hwang}
\affiliation{Materials Sciences Division,
Lawrence Berkeley National Laboratory, Berkeley, CA 94720, USA}

\author{Jack Deslippe}
\affiliation{Department of Physics, University of California,
Berkeley, CA 94720, USA}
\affiliation{Materials Sciences Division,
Lawrence Berkeley National Laboratory, Berkeley, CA 94720, USA}

\author{Alexei V. Fedorov}
\affiliation{Advanced Light Source, Lawrence Berkeley National Laboratory, Berkeley, CA 94720, USA}

\author{Steven G. Louie}
\affiliation{Department of Physics, University of California,
Berkeley, CA 94720, USA}
\affiliation{Materials Sciences Division,
Lawrence Berkeley National Laboratory, Berkeley, CA 94720, USA}

\author{Alessandra Lanzara}
\affiliation{Department of Physics, University of California,
Berkeley, CA 94720, USA}
\affiliation{Materials Sciences Division,
Lawrence Berkeley National Laboratory, Berkeley, CA 94720, USA}


\begin{abstract}
The Landau-Fermi liquid picture for quasiparticles assumes that charge carriers are dressed by many-body interactions, forming one of the fundamental theories of solids. Whether this picture still holds for a semimetal like graphene at the neutrality point, i.e., when the chemical potential coincides with the Dirac point energy, is one of the long-standing puzzles in this field.  Here we present such a study in quasi-freestanding graphene by using high-resolution angle-resolved photoemission spectroscopy. We see the electron-electron and electron-phonon interactions go through substantial changes when the semimetallic regime is approached, including renormalizations due to strong electron-electron interactions with similarities to marginal Fermi liquid behavior. These findings set a new benchmark in our understanding of many-body physics in graphene and a variety of novel materials with Dirac fermions.
\end{abstract}

\maketitle

The many-body interactions in graphene at the charge neutrality point, where the Fermi surface of graphene is pointlike, are expected to differ from those of an ordinary metal.  Due to the long-range nature of the electron-electron interaction, logarithmic velocity renormalizations and a linear imaginary self-energy (Im$\Sigma$) are expected by theoretical calculations, which differ from the case of a normal Fermi liquid \cite{CastroNeto,Gonzalez1,DasSarma,Trevisanutto,Polini,Profumo,Varma}.
The electron-phonon interaction also differs from that of a normal metal at the charge neutrality point, in that a finite self-energy is expected even in the absence of the ordinary electron-phonon ``kink'' \cite{MatteoCalandra,CohenLouieElPh}.  But despite the recent flurry of theoretical works, there has been no experimental demonstration of the self-energy in graphene at the neutrality point.

Angle resolved photoemission spectroscopy (ARPES) is the ideal tool to probe electron dynamics and many-body interaction in graphene \cite{himpselbook} as it can directly access the quasiparticle self-energy. 
So far however, most experimental studies of self-energy are on metallic graphene, i.e., when the chemical potential is far away from the neutrality point. In this regime many-body interactions are substantially screened \cite{Bostwick} and can still be overall described within the Fermi liquid picture.   
Although these samples can be successfully hole doped to bring the Fermi level closer to the Dirac point, due to the extreme sensitivity of graphene electronic structure to doping and/or disorder \cite{ChargedImpurityScattering}, the resulting spectra are much broader than the as-grown samples \cite{ZhouNo2,holedopinggoldbismuth,exfoliatedarpes} making it difficult to extract many-body information.
Recently it has been shown that a freestanding monolayer graphene film can be found on top of a bulk graphite system, and that the monolayer is decoupled and suspended over the entire substrate \cite{Andrei}.  Moreover it has been shown that its properties are nearly identical to a fully suspended graphene sheet \cite{Andrei}.
An almost identical system can be found when graphene is grown on the carbon face of SiC, SiC($000\overline{1}$),\cite{BergerScience}.  Due to the differing rotational orientations of the adjacent graphene layers with respect to the substrate, layers can behave like isolated graphene sheets\cite{BergerScience,MillerScience,Conrad}. 

Here we present a systematic study of the self-energy of quasifreestanding graphene, grown on the carbon face of SiC.  By comparing the experimental dispersions of quasifreestanding samples to theoretical calculations and highly-doped samples, we show that both the electron-electron and electron-phonon interactions go through substantial changes when the semimetallic regime is approached.  The electron-electron interaction exhibits logarithmic renormalizations in the band dispersions, changing the overall curvature of the band dispersions from the non-interacting or highly doped cases, and allowing us to extract the effective dielectric constant or fine structure constant $\alpha$; while the imaginary self-energy is roughly linear in energy.  We also show that the features of the electron-phonon interaction diminish significantly as the neutrality point is approached.

\section{Results and Discussion}

Fig. 1A shows slices of ARPES intensity through the Dirac cone of monolayer graphene, at the Brillouin zone corner, K.  This conical dispersion with a pointlike Fermi surface is the typical feature of Dirac fermions in undoped monolayer graphene \cite{Conrad,NovoselovNature}.
The monolayer nature of this sample is also supported by the single vertical $k_z$ vs $k_y$ dispersion shown in panel b.  If AB-stacked graphene were present, this dispersion would have a number of bands determined by the number of AB-stacked multilayers present, with band intensity that oscillates as a function of $k_z$ \cite{OhtaPRL}.  
To extract quantitative information about the doping level, in panel c we show the dispersion obtained by fitting the peak positions of the momentum distribution curves (MDCs), the intensity at constant energy as a function of momentum.  
From the separation between the $\pi$ bands at E$_F$ we conclude that the as-grown sample is slightly hole-doped with n = $8\times10^{10}$ cm$^{-2}$, which is comparable in both sign and magnitude to that of exfoliated graphene \cite{Andrei} and further supports the decoupled nature of this sample.  
From the slope of the fitted dispersion we can also extract the Fermi velocity, $v_F=1.10\times10^6 m/s$, which matches the value measured by scanning tunneling microscopy \cite{MillerScience}.

\begin{figure*}
\includegraphics[width=8.5 cm] {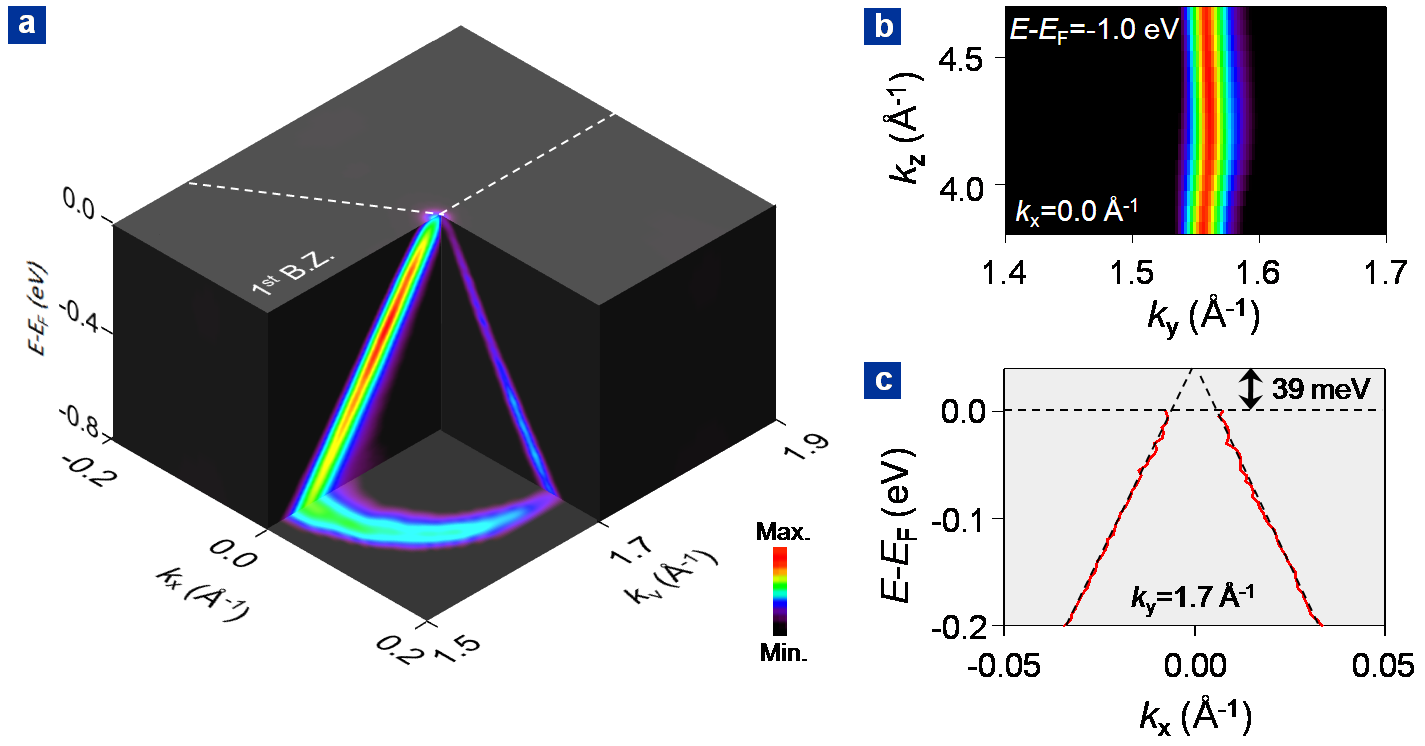}
\caption{The band structure of monolayer graphene.  (a) ARPES spectra of monolayer graphene, showing several slices through the Dirac cone of monolayer graphene.  (b) The $k_{z}$-dependence of photoemission intensity shows a single line, signature of monolayer graphene. (c) The red solid line shows the peak positions of the MDC fits, and the dotted red line represents the linear fit of the dispersion. The position of the Dirac point is slightly above the Fermi level.}
\end{figure*}

A close inspection of the electronic dispersion given in Fig. 2A (red solid line), shows that there is a significant departure from a linear bare band (dotted red line), most noticeable in the vicinity of E-E$_F$ = 800 meV.  This curvature can also be seen in the inset of panel a, where the Re$\Sigma$, defined as the difference between the bare band and the experimental peak position, is shown in the vicinity of the anomaly.
This anomalous curvature is most likely due to a many-body effect, such as the electron-electron interaction, rather than a bare band effect.  Indeed, both the LDA band curvature (dashed green line in the same panel) for undoped graphene and the valence band of highly doped graphene \cite{Bostwick} exhibit the opposite curvature to the experimental valence band reported here.  Among the many-body interactions, the electron-electron interaction is the most likely candidate: the electron-phonon coupling should affect the dispersion at a much smaller energy (the phonon energy) and should have the opposite curvature as we observe here \cite{MatteoCalandra,ZhouKink, CohenLouieElPh} (see Fig. 4), with decreasing velocity at the Fermi level; similarly, the electron-plasmon interaction is negligible for neutral graphene, and should also have the opposite curvature in any case \cite{Polini}.  The electron-electron interaction therefore appears to be the source for the anomaly.

\begin{figure*}
\includegraphics[width=8.5 cm] {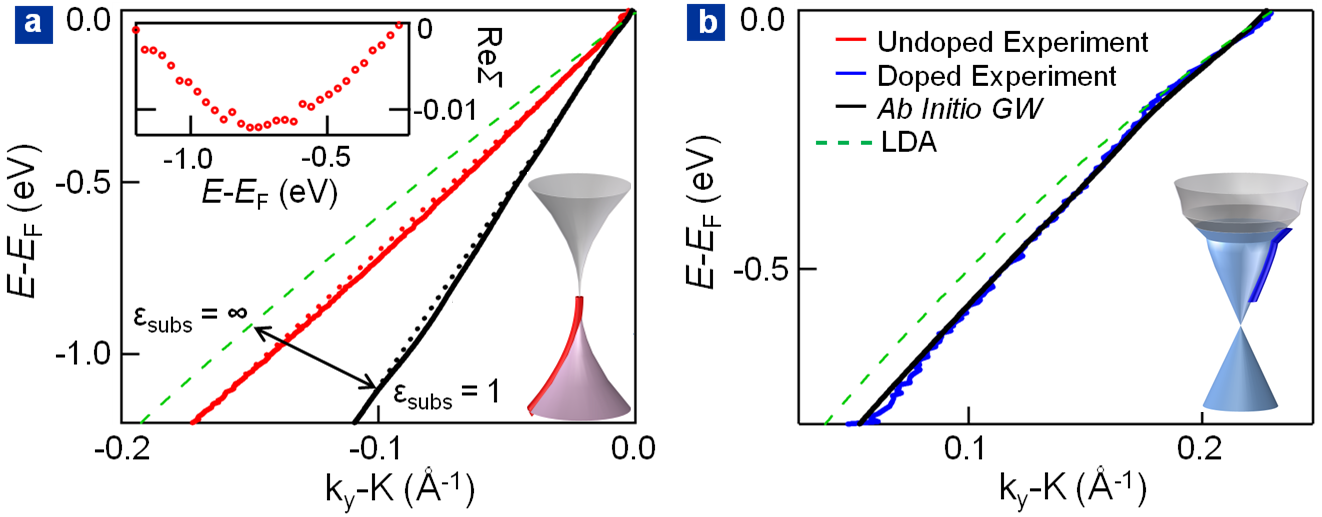}
\caption{Self-energy effect of the electron-electron interaction.  (a) Extracted dispersions from as-grown C-face graphene (red), and from \textit{ab initio GW} plus el-ph calculations including el-ph and el-el interactions in suspended graphene (black).  Straight red and black dashed lines are given as guides to the eye, which intersect the dispersions at -1.2 eV and -0.2 eV, to show how the experimental dispersions (solid lines) deviate from linearity (straight dotted lines).  The LDA (local density approximation) band is given as a green dashed line. Inset: Deviation of the experimental dispersion from linearity (red dotted line subtracted from the solid red experimental dispersion).  (b) Extracted dispersions for highly doped graphene (blue). Doping is obtained by K-deposition and the resulting charge density is more than three orders of magnitude higher than undoped graphene. Black solid line shows the \textit{ab initio GW} plus el-ph calculated dispersion.  The LDA band is shown by the green dashed line.}
\end{figure*}

This can be further confirmed by comparison between the experimental band and \textit{ab initio} calculations for freestanding graphene, performed within the \textit{GW} approximation \cite{CohenLouieElPhElEl} (see supplementary information), also shown in Fig. 2A.  The curvature of the electronic bandstructure is very similar when the electron-electron interaction is taken into account, as also seen in Ref. \cite{CohenLouieElPhElEl}.  In contrast, Ref. \cite{CohenLouieElPh} considers only the electron-phonon interaction and does not predict this effect; as stated above, the electron-phonon interaction causes the dispersion to curve in the opposite direction, with decreasing velocity at the Fermi level.  In summary Fig. 2 reveals our primary result: that the origin of this finite curvature is due to the electron-electron interaction and results in an increase of the band velocity as the Dirac point is approached.

Another interesting observation that follows from the comparison between the experimental and the theoretical dispersion in panel 2a is the decrease of the velocity, i.e., the slope of the dispersion, with increasing dielectric screening.  The LDA band, corresponding to infinite dielectric screening, has the smallest band velocity, while the \textit{ab initio} data corresponds to a fully suspended graphene sheet and has the largest velocity.  The experimental band velocity lies between these two extremes due to the finite dielectric screening of the graphitic substrate (the finite doping of the experimental data may also play a role).  Values of the band velocities near the Fermi level are listed on the left side of table 1.

\begin{table}
\caption{Summary of band velocities (m/s)} 
\centering     
\begin{tabular}{@{\vrule height 10.5pt depth4pt  width0pt}c cccc}  
\hline\hline                        
   &\multicolumn{2}{c}{Undoped} &\multicolumn{2}{c}{Highly Doped}\\ [0.5ex]
\hline
Dispersion & V$_{Fermi}$ & V$_{High}$ & V$_{Fermi}$ & V$_{High}$\\
\hline                
LDA   & 0.85$\times$10$^6$ & 0.89$\times$10$^6$ & 0.51$\times$10$^6$ & 0.61$\times$10$^6$ \\  
Experiment & 1.10$\times$10$^6$ & 1.13$\times$10$^6$ &  0.49$\times$10$^6$ & 0.77$\times$10$^6$ \\
\textit{Ab Initio} GW plus el-ph     & 1.48$\times$10$^6$ & 1.80$\times$10$^6$ &  0.58$\times$10$^6$ & 0.74$\times$10$^6$ \\[1ex] 
\hline                          
\end{tabular}
 
\begin{tabular}{@{\vrule height 10.5pt depth4pt  width0pt}l}
V$_{High}$ is the band velocity at approximately 300meV,\\
slightly higher binding energy than the phonon energy.
\end{tabular}
\end{table}

In contrast to undoped graphene (panel a) the effect of the electron-electron interaction and dielectric screening is reduced for highly doped graphene (panel b), resulting in a better overall agreement with the ab intio calculations (a summary of band velocities can be found in table 1).  The extent to which this might be true can be examined by considering the inverse screening length for doped graphene, which is given by $q_s = 4 \alpha k_F$ \cite{Kotov}.  Our as-grown sample has a Fermi vector of $k_F = 0.005$ \AA$^{-1}$, while the highly-doped sample has a Fermi vector of $k_F = 0.23$ \AA$^{-1}$.  Using our experimentally determined value of $\alpha = 0.4$ (discussed below), we find length scales corresponding to 790 \AA\ and 17 \AA\ for the as-grown and highly-doped samples, respectively.  In addition to verifying the validity of the \textit{ab initio} calculations, a close inspection of the highly-doped dispersions shows that they curve in the opposite direction as the as-grown sample.  These results clearly indicate that the many-body physics of freestanding graphene near the charge-neutrality point is dramatically different from highly-doped graphene.

To further investigate the implications of these findings, in Fig. 3a we consider Hartree-Fock and renormalization group corrections to the self-energy to account for the high energy band curvature observed in Fig. 2a.
It was proposed that, in contrast to the linear real self-energy of a Fermi liquid where quasiparticles decay via intraband transitions, when graphene is near the charge neutrality point, the electron-electron interaction leads to a logarithmic correction of the self-energy \cite{CastroNeto, Gonzalez1994}. This logarithmic self-energy can be approximated as Re$\Sigma_{ee}(k)=\frac{\alpha}{4}v_F^o |k-k_F|\ln(|\frac{k_c}{k_D-k}|)$, where $k_F$ is the Fermi wavenumber along the $\Gamma-K$ direction, $k_D=1.703$ \AA$^{-1}$ is the Dirac point wavenumber, $|k_D-k_F|=0.005$ \AA$^{-1}$, $k_c=1.703$ \AA$^{-1}$ is a cutoff, and the fit parameter $\alpha$ is the dimensionless fine structure constant of graphene, $\alpha=\frac{e^2}{\epsilon v_F^o}$.  Consequently, the dispersion relation is modified as $E(k)=-v_F^o |k-k_D| + \Sigma_{ee}(k)$.  We obtain a bare band velocity of $v_F^o=(0.86\pm0.02)\times10^6$ m/s (which is smaller than the experimental Fermi velocity of 1.10$\times10^6$ m/s) and an effective dielectric constant $\epsilon=6.4\pm0.1$ (or $\alpha=0.40\pm0.01$).
The two parameter fit was performed over the energy range [-1.3 eV,-0.4 eV] to avoid contributions from the electron-phonon kink, in addition to the constraint that the band passes through $E(k=k_F)=0$, to account for the low binding energy deviations in the vicinity of the phonon kink (which will be discussed later), and some finite-resolution effects near the Fermi level. The very good agreement between the logarithmic fit and the measured dispersion (panel 3a) points to a clear signature of renormalizations due to electronic correlations. 
This self-energy takes the same form as that of a marginal Fermi liquid, although higher order corrections might lead to small deviation from this picture \cite{Kotov}.
Note that, although obscured by the presence of the phonon kink, electron-electron renormalizations remain apparent at low binding energy: the experimental band velocity remains larger than that of the LDA band, and the electron-electron interaction is the only interaction known to result in an increase in the Fermi velocity. 

\begin{figure*}
\includegraphics[width=8.5 cm] {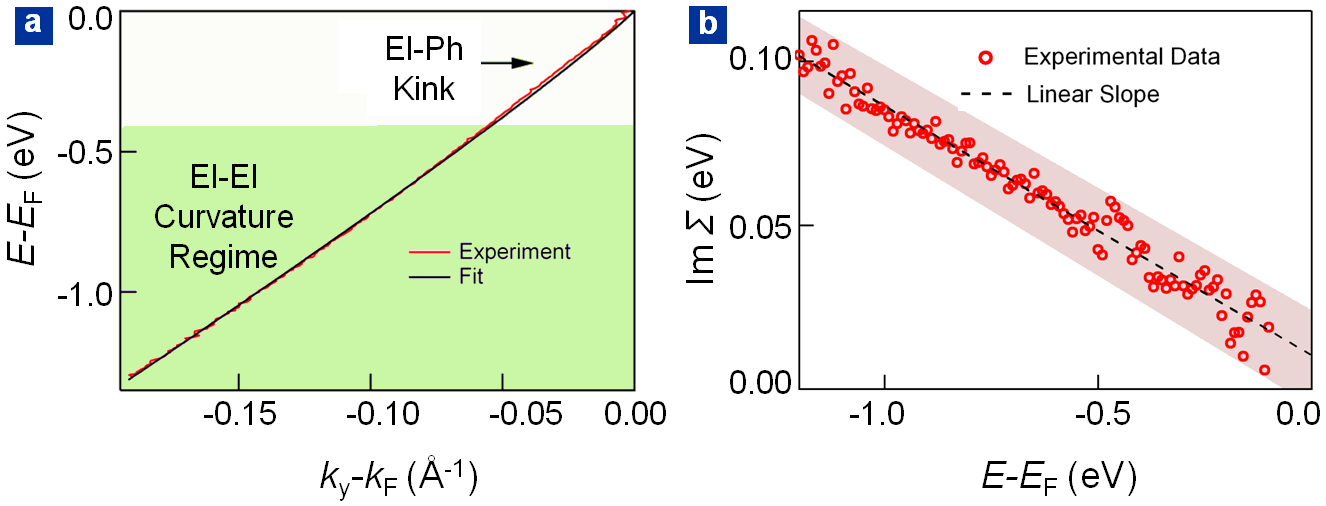}
\caption{Comparison of the experimental dispersions to Hartree-Fock and renormalization group self-energy.  (a) The experimental dispersion (red) and theoretical fit (black) given by the electron-electron self-energy correction to the bare band.  The fit is given by $E(k_y)=5.63 (k_y-k_F)-0.563 (k_y-k_F) \ln(|\frac{k_y-k_D}{1.7}|)$.  (b) Im$\Sigma$ extracted from Voigt fits to MDCs.  The dotted line is a guide to the eye.}
\end{figure*}

The extracted value for the dielectric screening falls within the large range of values reported in the literature, which are influenced by the choice of substrate.
These values range from $\epsilon=15$ for a freestanding layer\cite{Abbamonte}, to $\epsilon=4.4$ for monolayer graphene on SiC, although this value increases greatly in the presence of an interfacial carbon layer\cite{Plasmaron}.  Theoretical calculations typically discuss a dielectric constant of the form $\epsilon = (\epsilon_{substrate} + \epsilon_{vacuum})/2 = (\epsilon_{substrate} + 1)/2$, with an additional term for screening from the graphene film itself: calculations expect $\epsilon=4$ for fully suspended graphene\cite{DASSARMA1}.  A comparison between our data and these calculations lead us to expect that the graphitic substrate has a dielectric constant of roughly $\epsilon=7$.  This value is of the same order of magnitude as previous studies, although the real dielectric constant of graphite is a source of controversy with measured values ranging from 2 to 13 \cite{Jellison}. 

The presence of a logarithmic term in the self-energy, and hence the departure from the Fermi liquid picture, should also be manifested in the Im$\Sigma$, predicted to scale linearly with energy \cite{Gonzalez1}.  
This can be observed in Fig. 3b, where a linear Im$\Sigma$ is observed over a large energy window.  The Im$\Sigma$ is extracted from the experimental data using the standard fitting procedure by Voigt lineshapes of the raw MDCs \cite{voigt}, where the Voigt functions are defined as the convolution between a Gaussian, to account for the resolution broadening, and a Lorentzian, the expected MDC lineshapes in the presence of a $k$-independent spectral function. 
This fitting procedure allows one to isolate the intrinsic contribution to the self-energy, by separating extrinsic parameters such as experimental resolution and sample inhomogeneities \cite{voigt}.
Note that the linear self-energy extends up to more than 1 eV, as expected in the case of conical dispersion \cite{PeresPRB06}, where the energy range of validity of the logarithmic renormalizations roughly coincides with the energy where the band structure loses its linear dispersion relation \cite{Xu}.  However, we should also note that the electron-phonon interaction adds a significant contribution to the self-energy, which in undoped graphene is predicted to increase in binding energy in a similar manner \cite{CohenLouieElPh}.

These findings are in line with infrared spectroscopy studies, where electronic correlation effects were reported to modify the Fermi velocity \cite{Basov}.  It also supports the view that electronic compressibility measurements \cite{Yacoby} can be explained in terms of electronic interactions, as discussed by Sheehy and Schmalian \cite{Schmalian}.
  
In Fig. 4 we focus on the low energy regime, in the range of 150-200 meV from E$_F$, where another deviation of the dispersion relation from linearity is observed.  This deviation, also known as kink, has been widely studied for highly doped graphene and is a signature of electron-phonon interaction \cite{MatteoCalandra,ZhouKink}.  What is surprising about these data is that, while for doped graphene the effect of the electron-phonon interaction is strongly manifested in the dispersion (panel b) and in both the Re$\Sigma$ and Im$\Sigma$ (panels c and d respectively) in the form of a peak and a step at the phonon energy respectively, it is strongly reduced for neutral graphene, where only a tiny deviation from linearity can be discerned (panel a and c).  This translates into a featureless Im$\Sigma$ (panel d).

\begin{figure*}
\includegraphics[width=8.5 cm] {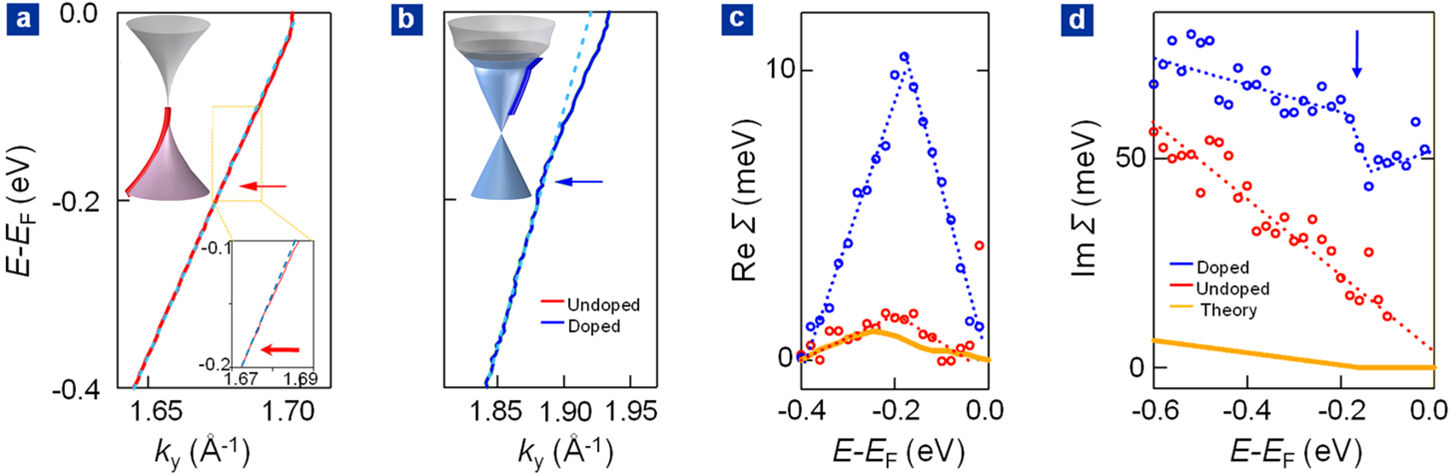}
\caption{Doping dependence of the electron-phonon interaction.  The dashed lines are guides to the eye.  (a) Dispersion (solid red line) extracted along the $\Gamma$-K direction for the as-grown graphene near zero doping.  A small deviation from linearity (light blue dashed line) is indicated by the arrow.  (b) Dispersion (solid blue line) extracted from the $\Gamma$-K direction of highly-doped graphene.  A kink due to electron-phonon coupling is indicated by the horizontal arrow (the light blue dashed line is a guide to the eye). (c) Comparison of Re$\Sigma$ for doped graphene (blue), undoped (red) graphene, and el-ph calculation (orange).  To approximate the bare band, a linear offset has been subtracted, forcing each curve to intersect zero energy at -0.4 eV and 0 eV.  (d) Comparison of Im$\Sigma$ for doped graphene (blue), undoped graphene (red), and electron-phonon calculation (orange).  For ease of viewing, an offset was removed from the highly-doped data.}
\end{figure*}

This strong reduction of the kink for neutral graphene is well reproduced by theoretical calculations of the electron-phonon self-energy\cite{CohenLouieElPh} where electron-electron interactions (and others) are not included (panels c and d).  On the other hand, the slopes of the theoretical and experimental self-energies differ greatly when only the electron-phonon self-energy is considered, since the electron-phonon interaction is only one component of the total self-energy.
It is important to note that the absence of a feature in the self-energy does not necessarily imply that the self-energy is zero.  Since each of these interactions have self-energy components that increase with binding energy, we can only compare the distinct features of these interactions.  In the case of the electron-phonon interaction, undoped graphene has a finite self-energy, but no step in the Im$\Sigma$ at the phonon energy, while the kink in the Re$\Sigma$ is diminished as well \cite{CohenLouieElPh}.  These effects are clearly indicated in our data, and differ from the behavior of ordinary metals.

\section{Conclusion}

In summary, we have demonstrated that many-body interactions in monolayer graphene are substantially modified when the neutrality point is approached, with a departure of the self-energy from ordinary metallic systems. The electron-electron interaction results in novel renormalizations, while the effects of the electron-phonon interaction diminish as a function of decreasing doping.  These findings are the result of the point-like Fermi surface and linear bands of monolayer graphene, which set the many-body interactions in graphene apart from those of ordinary metals.

\section{Materials}
The graphene layers were grown in a closed rf induction furnace at a temperature of 1550$^\circ$C (see Ref. \cite{HassReview} for details).  The sample was annealed to 1000$^\circ$C in ultra-high vacuum (UHV) prior to measurement.  Our ARPES investigation was performed at beamline 12.0.1.1 at the Advanced Light Source, at a pressure better than 3$\times$10$^{-11}$ torr, with a sample temperature of 15$^\circ$K, photon energies of 42 eV and 50 eV.  All of the ARPES measurements in this study were performed on a single rotational domain of a single graphene sample.  The lateral extent of this rotational domain is less than 30 $\mu$m$\times$80$\mu$m, the approximate dimensions of the ARPES beamspot.  This sample was then electron-doped \textit{in situ} by potassium deposition with an SAES Getters alkali metal dispenser, allowing us to study the same position on the sample for both the as-grown and highly-doped cases.  \textit{Ab initio} calculations including the contributions of electron-electron and electron-phonon interactions to the self-energy were performed as in the supplementary information.

\section{acknowledgments}
We are greatly indebeted to Joerg Schmalian, Sudip Chackravarty, and Antonio Castro Neto for englightening discussions.  We also acknowledge Walt de Heer and Claire Berger (Georgia Institute of Technology, Atlanta, GA) for providing us with the high quality graphene samples which have made these studies possible. 
This work was supported by the Director, Office of Science, Office of Basic Energy Sciences, Materials Sciences and Engineering Division, of the U.S. Department of Energy under Contract No. DE-AC02-05CH11231.  C.H.P. was supported by ONR MURI under Grant No. N00014-09-1-1066.  The sample growth and part of the ARPES measurements were supported by the MRSEC (Materials Research Science and Engineering Centers) under National Science Foundation Grant No. DMR-0820382.  Parts of the computations were done with theoretical developments and codes supported by National Science Foundation Grant No. DMR10-1006184.

Correspondence and requests for materials should be addressed to Alanzara@lbl.gov.

\end{document}